\documentstyle[12pt]{article}

\input{epsf}

\renewcommand{\baselinestretch}{1.1}

\def\bsgam{b\rightarrow s\gamma}
\def\as{\alpha_s}
\def\be{\begin{equation}}
\def\ee{\end{equation}}
\def\bea{\begin{eqnarray}}
\def\eea{\end{eqnarray}}

\def\sign{{\sf sign}}

\def\uln{\underline}

\begin{document}

\begin{titlepage}

\makebox{ }
\vspace{ -1in}

\begin{tabbing}
xxxxxxxxxxxxxxxxxxxxxxxxxxxxxxxxxxxxxxxxxxxxxxxxxxxx\=xxxxxxxxxx\kill
 \> IUHET-376 \\ 
 \> OHSTPY-HEP-T-97-023 
\end{tabbing}


\vskip 0.5in

\begin{center}

{\large \bf 
$b\rightarrow s\gamma$ with Large tan$\beta$ in MSSM Analysis 
Constrained by a Realistic SO(10) Model} 

\vskip 0.4in

Tom\'{a}\v{s} Bla\v{z}ek$^{a*}$ and
Stuart Raby$^{b\dagger}$ \\

\vskip 0.2in

$^a${\em Department of Physics, Indiana University,}\\
    {\em Swain Hall West 117, Bloomington, IN 47405, USA} \\
\vskip 0.1in
$^b${\em Department of Physics, The Ohio State University,}\\
    {\em 174 W. 18th Ave., Columbus, OH 43210, USA} \\
\vskip 0.2in

December 3, 1997 \\

\end{center}

\vskip 0.7in

\begin{abstract}

Study of the MSSM in large tan$\beta$ regime has to include correlations 
between the constraints presented by the low energy values of the $b$ quark 
mass and BR($\bsgam$). Both quantities receive SUSY contributions enhanced 
by tan$\beta$ and have a major impact on the MSSM analysis. Here we 
summarize the results of such a study constrained by a complete SO(10) model. 
The dominant effects to the analysis come from third generation
Yukawa couplings and a GUT threshold to $\as$.
We show that a small negative GUT correction 
to $\as$ accommodates $\as(M_Z)\le 0.118$ and $\delta m_b^{SUSY} > 0$.
The latter quantity being positive then opens up two options to fit the 
measured rate for $\bsgam$. The two distinct fits differ by the overall sign 
of the amplitude for this process. They work equally well in complementary 
regions of the allowed SUSY parameter space. We show plots of the partial 
contributions to the coefficient $C_7(M_Z)$ in the ($m_0,M_{1/2}$) plane 
in each of these best fits. We conclude that an attractive SO(10)-derived 
regime of the MSSM remains a viable option.

\end{abstract}

\vskip 0.5in


\vskip 0.1in

$^*${\footnotesize On leave of absence from 
the Dept. of Theoretical Physics, Comenius Univ., Bratislava, Slovakia;\\ 
\makebox[2.1em]{ }blazek@gluon2.physics.indiana.edu} \\

$^\dagger${\footnotesize raby@pacific.mps.ohio-state.edu}

\end{titlepage}


\renewcommand{\thepage}{\arabic{page}}
\setcounter{page}{1}
\renewcommand{\baselinestretch}{1.3}

\section{Introduction}

\indent
The inclusive $\bsgam$ decay has attracted a lot of attention in the past
and will undoubtfully attract at least as much attention in the future. 
On one side, it is an observed flavor changing neutral current (FCNC) 
process with the measured rate 
$BR(\bsgam)=(2.32\pm0.57\pm0.35)\;\times \; 10^{-4}$
announced by CLEO\cite{bsgam_exp_CLEO}, and 
$BR(\bsgam)=(3.38\pm0.74\pm0.85)\;\times \; 10^{-4}$
found most recently by the ALEPH Coll.\cite{bsgam_exp_A},
and we can expect that the statistical uncertainty of these measurements
will be reduced in the near future.
On the other side, the Standard Model (SM) prediction with the 
next-to-leading order QCD corrections included has been calculated as 
$BR(\bsgam)=(3.48\pm0.31)\;\times \; 10^{-4}$
\cite{NLO_SM}. However, since forbidden at tree level 
it lets the SM compete with the loop contributions from any new physics.
Theoretical analyses of this process, which assume minimal supersymmetric
(SUSY) extension of the SM and are constrained by unification,
naturally fall into two categories. Either tan$\beta$ (the 
ratio of the Higgs vacuum expectation values $\,<\!H^0_u\!>\!/\!<\!H^0_d\!>$ ) 
is considered to be low ({\em i.e.,} close to 1) or this parameter is large, 
of the order 50.

In the first case, both $\,<\!H^0_u\!>$ and $<\!H^0_d\!>$ are of the order 
of the electroweak scale and the hierarchy $M_b,M_\tau \ll M_t$ is triggered
by the smallness of the $b$ and $\tau$ Yukawa couplings $\lambda_b$ 
and $\lambda_\tau$. In effect, this regime of the Minimal Supersymmetric 
Standard Model (MSSM) has several simple features: radiative electroweak 
symmetry breaking is rather straightforward to impose, and the terms suppressed
by $\lambda_b$ and $\lambda_\tau$ can be neglected both in the renormalization 
group equations (RGEs) and in the analysis at the $Z$ scale. 
A drawback, if we can say so, is that when 
considering the MSSM as a low energy effective theory, there is no appealing 
unification dynamics requiring the hierarchy $\lambda_b$,$\lambda_\tau \ll 
\lambda_{top}$.

The large tan$\beta$ case, on the other hand, is very attractive when 
considered from the unification point of view. Using the MSSM RGEs one 
finds that this case offers an amazingly simple option
$\lambda_b =\lambda_\tau = \lambda_{top}$ at the unification scale, 
consistent with the minimal $SO(10)$ grand unified theories (GUTs).
The third generation mass hierarchy is, however, explained by 
$<\!H^0_d\!>\:\approx 3$-$4$GeV, the value which is much less than the 
electroweak scale - the generic scale of the other dimensionful parameters, 
including $<\!H^0_u\!>$. That makes the MSSM analysis with large tan$\beta$ 
challenging in more than just the symmetry breaking sector. For example, there 
are fermion masses which are proportional to $<\!H^0_d\!>$ at tree level, 
and there is no symmetry which would guarantee the same suppression in 
their radiative corrections. Contributions at one-loop may be and indeed 
are proportional to $<\!H^0_u\!>$. The same type of corrections are
then also induced to some CKM matrix elements, e.g. $V_{cb}$ or $V_{ub}$.
As a result, the tan$\beta$ enhanced corrections 
to such observables must be included into the MSSM analysis. 
Inclusive $\bsgam$ decay represents an even more sensitive probe if tan$\beta$
is large. As an FCNC process emerging only at a loop level 
its SM contribution turns out to be suppressed by 
$<\!H^0_d\!>$. On the other hand, the chargino-squark contributions 
are proportional to $<\!H^0_u\!>$ and may dominate the whole process. 
Thus the MSSM analysis becomes more powerful (and restrictive) 
than in the case with low tan$\beta$. 

In this work, we present the results of such a constrained global
analysis of the MSSM with large tan$\beta$ and pay special attention to 
the structure of the $\bsgam$ partial amplitudes. Our actual investigation 
has been performed within a complete $SO(10)$ model proposed 
by Lucas and Raby\cite{lucas}. The model 
(called model 4c), was analyzed in detail in \cite{ga_newPRD} 
and a very good agreement with low energy data (including fermion masses 
and mixings) was found in a large region of the SUSY parameter space. 
Although the primary goal of the analysis was to discriminate among different
models of fermion masses, this model turned out to work so well, that 
the main constraints were the $b$ quark mass and the $BR(\bsgam)$. As a result 
we can regard the work as testing the MSSM with large tan$\beta$,
with the inclusion of a particular set of 3$\times$3 non-diagonal 
Yukawa matrices at the GUT scale. Thus 
we think that this letter naturally fits into the mosaic of many 
previous works on $\bsgam$ in the MSSM.
These works studied the $\bsgam$ process 
either with low (or moderate\footnote
{with no reference to the unification of the Yukawa couplings
}, 
i.e. less than about 10) tan$\beta$ only\cite{bsgam_theory_lowtb} 
or, if the large tan$\beta$ regime was also considered 
\cite{bsgam_theory_largetb}, then some of the important ingredients of our 
procedure were not taken into account. 
Most notably, we improve the previous studies {\em (i)} by taking into
account the tan$\beta$ enhanced SUSY threshold corrections to fermion masses
and mixings, and 
{\em (ii)} by introducing global analysis instead of scatter plots of 
random points in SUSY parameter space. Due to the latter improvement 
we can present the contour plots of the interesting quantities from the best 
fits and observe how these quantities are correlated.
Additional motivation for this paper came from the study \cite{deBoer}. This 
work has also introduced global analysis, with one-loop SUSY threshold
corrections properly included, and studied both the low and large tan$\beta$
regimes. However, this analysis assumed strict gauge unification with no
threshold correction 
to $\as$ at the GUT scale. As a result, it did not concentrate on the 
regime with rather low values of $\as(M_Z)\leq 0.118$ and positive 
SUSY corrections to $m_b$. Therefore, that work does not test the subspace 
in SUSY parameter space which we study in this analysis\footnote
{We have, in fact, observed 
identical features in the same 
subspace of the SUSY parameter 
space and have come to the 
same conclusions as in 
\cite{deBoer_Jerus} that the
subspace where $\delta m_b^{SUSY}<0$
appears to be excluded 
---
see the discussion on negative $\mu$ 
parameter in section 3 and in 
\cite{ga_newPRD}. (In our conventions, 
$\mu\!<\!0$ implies negative SUSY 
corrections to the $b$ quark mass 
and purely constructive interference 
among all leading MSSM 
contributions to the decay $\bsgam$.)
}.

Thus our main result is that in the regime of SUSY parameter space
where tan$\beta$ is large and the $b$ quark mass receives positive corrections
we find the analysis consistent with $t-b-\tau$ unification and 
the observed BR($\bsgam$), in contradiction to previous studies. 
Dominant effects come from the third generation Yukawa couplings and
the introduction of $\epsilon_3$, a GUT threshold to $\as$. In the best fits,
small negative $\epsilon_3$ is correlated with lower $\as(M_Z)$, 
and that in effect decreases $m_b$. 
We also find that the contribution of the inter-generational 
$\tilde{c}_L$-$\tilde{t}_L$ squark mixing to the $\bsgam$ amplitude
(tan$\beta$ enhanced compared to the SM contribution) is numerically 
significant. It is, however, model dependent. 
Note that it is neglected in the approximation of Barbieri-Giudice
\cite{bsgam_theory_largetb} which has been used in many of the succeeding
studies.

The main steps of our procedure are described 
in section 2. For completeness, the performance of model 4c in 
the global analysis is presented in section 3. Section 4 contains the results
of this work. It starts with comment on model sensitivity
and includes contour plots in the ($m_0,M_{1/2}$) plane 
of constant $BR(\bsgam)$ and various contributions to this process, 
extracted from our best fits. We show how 
the destructive interference among the $\bsgam$ partial amplitudes 
provides for two distinct fits, each of a very low $\chi^2$ in
rather complementary regions of the parameter space.
We also discuss phenomenological implications of our findings. 
Finally, conclusions in section 5 contain a brief summary of this work.

\section{Global Analysis}

\indent
Details of our numerical analysis are described in \cite{ga_newPRD}.
Here we summarize the main steps of our procedure relevant for 
this letter. We perform a global analysis in a strict top-down approach. 
We start with the following initial parameters: the scale of new physics 
$M_G$, unified gauge coupling $\alpha_G(M_G)$ 
\footnote
{$M_G$ is defined as the scale 
where the gauge couplings 
$\alpha_1$ and $\alpha_2$ 
are exactly equal within the 
one-loop GUT threshold 
corrections. By $\alpha_G$ 
we actually mean the value 
$\alpha_1(M_G)\equiv 
            \alpha_2(M_G)$
}, 
one-loop GUT threshold correction $\epsilon_3$ to $\alpha_3(M_G)$,
and A: the 33 element common to all three Yukawa matrices at $M_G$.
We assume supergravity induced SUSY breaking and neglect the effects of 
running between the Planck scale and the GUT scale.
The parameters of the SUSY sector, which are introduced at $M_G$, include 
a common gaugino mass $M_{1/2}$, common scalar mass $m_0$ of squark and 
slepton mass matrices, scalar Higgs mass parameters $m_{H_d}$ 
and $m_{H_u}$ and a universal dimensionful trilinear coupling $A_0$.
For simplicity, the $\mu$ parameter and its SUSY breaking bilinear partner
$B$ are introduced at the $Z$ scale, since they are renormalized 
multiplicatively and do not enter the RGEs of the other parameters.
In the actual model 4c analysis, the structure of the Yukawa matrices at $M_G$ 
is built up using six more dimensionless parameters. As explained in 
section 4, these are small numbers and their effects decouple 
from the observables related to the gauge and SUSY sectors and from the
masses of the third generation fermions.

We use the two-loop RGEs for the gauge and Yukawa couplings, and the 
one-loop RGEs for the MSSM dimensionful parameters to run down to the 
$Z$ scale. At selected points, we check that the full two-loop RGEs 
of the MSSM \cite{mv} yield the same results. At the $Z$ scale, the
MSSM is matched to the effective theory consisting of QCD and 
electromagnetism, leaving out the SM as an effective theory. 
Within the MSSM, we implement the 
effective potential method 
of \cite{vev} 
(at one loop, with all one-loop threshold effects included)
to obtain the fit values of $v$ and tan$\beta$. 
These quantities are not directly restricted. (Neither is tan$\beta$ among
the free initial parameters.) Electroweak symmetry breaking is established 
implicitly in the process of the $\chi^2$ minimization by the evaluation 
of the fermion and gauge boson masses. The latter are calculated with the 
full MSSM one-loop corrections included. We also compute all the threshold 
corrections, proportional to tan$\beta$, to fermion masses and CKM matrix 
elements, following \cite{CKM_corr}. Masses of the Higgs particles 
are calculated at the two-loop level as in \cite{vev}, and masses of the SUSY 
particles are left at their respective tree level values.
The tree level squark and slepton masses are constrained to be greater than 
30 GeV. This is below the experimental limit but when they are that light, 
we count on substantial enhancements at one loop \cite{damien_SUSY_spectra}. 
Chargino and neutralino masses receive very small one-loop corrections and 
so we restrict their tree masses by the respective LEP limits.
As we match the MSSM to the effective theory below $M_Z$, the 
threshold corrections to the gauge couplings are calculated following 
\cite{chank}, with the exception that the SUSY vertex and box corrections 
to $\Delta r$ are neglected. In our approach, $\Delta r$ serves to derive
the theoretical value of $G_\mu$. When below the $Z$ scale, we use the 
three-loop QCD and one-loop QED RGEs \cite{RGEs_below_Z} to evaluate the 
quark and lepton masses. 

\protect
\begin{table}[t]
$$ 
\begin{array}{|r|c|c|c||r|c|c|c|}
\hline
\multicolumn{2}{|c|}{\rm Observable} & {\rm Central}& \sigma &
\multicolumn{2}{ c|}{\rm Observable} & {\rm Central}& \sigma \\
\multicolumn{2}{|c|}{  }             & 
\makebox[1.7cm]{\rm value}  & \makebox[1.7cm]{} &
\multicolumn{2}{ c|}{  }             & 
\makebox[1.7cm]{\rm value}  & \makebox[1.7cm]{} \\
\hline
 1. & M_Z              &  91.186      & \uln{0.46}      &
11. & M_b - M_c       &    3.4        &  0.2       \\
 2. & M_W              &  80.356      & \uln{0.40}      &
12. & m_s             &  180          & 50         \\
 3. & G_{\mu}    &  1.166\cdot 10^{-5} &\uln{1.2\cdot 10^{-7}} &
13. & m_d/m_s         &  0.05         &  0.015     \\
 4. & \alpha^{-1}     &  137.04       & \uln{0.69}      &
14. & Q^{-2}          &  0.00203      &  0.00020   \\
 5. & \alpha_s(M_Z)   &  0.118        &  0.005     &
15. & M_{\mu}         & 105.66        & \uln{0.53}      \\
\cline{1-4}
 6. & M_t             &  175.0        &  6.0       &
16. & M_e             &  0.5110       & \uln{0.0026}    \\
\cline{5-8}
 7. & m_b(M_b)        &    4.26       &  0.11      &
17. & V_{us}          &  0.2205       &  0.0026    \\
 8. & M_{\tau}        &  1.777        & \uln{0.0089}    &
18. & V_{cb}          &  0.0392       &  0.003     \\
\cline{1-4}
 9. & \rho_{new} & -0.6 \cdot 10^{-3}  & 2.6\cdot 10^{-3}   &
19. & V_{ub}/V_{cb}   &  0.08         &  0.02      \\
10. & B(b \rightarrow s \gamma) &  2.32\cdot 10^{-4} &  0.92\cdot 10^{-4}  &
20. & \hat B_K        &  0.8          &  0.1       \\
\hline
\end{array}
$$
\caption{Experimental observables of the global analysis.}
\label{t_obs}
\end{table}

Our $\chi^2$ function is calculated based on the low energy data (observables 
and their corresponding errors) listed in table \ref{t_obs}. Our analysis was
originally designed to test GUT models, so from the point of view of testing 
the MSSM the data in table \ref{t_obs} are divided into two groups,
corresponding to observables 1-10, and 11-20 respectively. Five 
observables in the gauge sector ($M_Z,\,M_W,\,G_\mu,\,\alpha$ and $\as$), 
masses of the third generation fermions, $\rho_{new}$
\footnote
{This is the contribution of physics 
beyond the SM to the $\rho$ parameter. 
} 
and $BR(\bsgam)$ are typically chosen to test the MSSM constrained by 
unification. 
The other ten observables corresponding to six light fermion masses  
and four independent parameters of the CKM matrix have been included
in the analysis but they do not significantly affect our results
(see introduction to section 4 where we discuss model dependence).

Note that seven out of the twenty observables have the estimated 
theoretical uncertainties dominating over the experimental ones
(their respective $\sigma$'s are underlined in table \ref{t_obs}). 
These theoretical uncertainties
represent conservative estimates of the errors (0.5\% for six out of the 
seven, and 1\% for $G_\mu$ to compensate for SUSY boxes and vertices which 
are neglected in the computation of $\Delta r$) generated by our numerical
procedure. In addition, note that we have also introduced a conservative
error on $\as(M_Z)$ \cite{Burrows} and added the CLEO errors for the 
$BR(\bsgam)$ linearly.

\subsection{Calculation of $\bsgam$}

The MSSM amplitude for the transition $\bsgam$ is calculated following 
\cite{bbmr} at the threshold $M_Z$.  
The effective Hamiltonian method, summarized in \cite{buras}, is used 
below $M_Z$. In particular, the amplitude at $M_Z$ 
is matched to the Wilson coefficient $C_7(M_Z)$ in the effective 
hamiltonian
\be
    H_{eff} = - \frac{4G_F}{\sqrt{2}}V^*_{ts}V_{tb}\;
              \sum_{i=1}^8 C_i(\mu)O_i(\mu)\;.
\label{H_eff}
\ee
Following the conventions of ref.\cite{buras} the magnetic dipole 
operator reads
\be
    O_7     = \frac{e}{16\pi^2}\,m_b\, (\bar{s}_L\sigma^{\mu\nu}b_R)
                                     \,F_{\mu\nu}\;,
\label{O_7}
\ee
with $\sigma^{\mu\nu}=i/2\,[\gamma_\mu,\gamma_\nu]$.
The branching ratio is computed from the formula
\be
 BR(\bsgam) = \frac{|V^*_{ts}V_{tb}|^2}{|V_{cb}|^2}\;
              \frac{6\alpha}{\pi g(M_c/M_b)}\; |C_7^{eff}(\mu_b)|^2\;
              BR(b\rightarrow ce\bar{\nu})\;,
\label{BR}
\ee
where $BR(b\rightarrow ce\bar{\nu})=0.104$ \cite{PDG}, $\alpha=1/132.5$ 
and the phase-space function for the semileptonic decay 
$g(z)=1-8z^2+8z^6-z^8-24z^4\log z$. The effective coefficient
\be
  C_7^{eff}(\mu) = \eta^{16\over23}C_7(M_Z) 
                  +{8\over 3}\:(\eta^{14\over23} - \eta^{16\over23})\:C_8(M_Z)
                  +C_2(M_Z)\,\sum_{i=1}^8h_i\eta^{a_i}\,
\label{C_7_eff}
\ee
comes from operator mixing in the leading log approximation, 
with $\eta= \as(M_Z)/\as(\mu)$. The numbers $h_i$ and $a_i$ are given in 
\cite{buras}. For the CKM matrix elements and quark masses, the fit values 
of the model 4c analysis are consistently used in the formulas above. 
The best fits yield their values very close to the ones quoted by Particle 
Data Group (PDG) \cite{PDG}. 
\footnote
{Since we do not choose some fixed values, 
we observe a moderate dependence 
of the $BR(\bsgam)$, within 10\%, 
on the particular set of the 
computed theoretical values for these 
quantities. For the most part, it is
due to the function g(z) which changes 
rather fast in the vicinity of 
$z\!\approx\! 0.3$.
}
The values of $\as(M_Z)$ are also taken from the GUT analysis. (The best 
fit values which are used to calculate $\eta$ in eq.(\ref{C_7_eff}) 
can be read out from the figures discussed in section \ref{ss_as}.)

In our analysis, we fix the low energy scale to be $\mu_b=4.7$GeV and do 
not study the scale dependence of the result. The scale dependence will be
reduced once the complete next-to-leading order calculation within the MSSM 
will be known. Because of the significant SUSY contributions to this process 
in large tan$\beta$ regime, we use the leading order calculation in our fits 
and only comment (in section \ref{ss_NLO}) on possible changes which may 
result from the next-to-leading order calculation in the full MSSM. 

\section{Model 4c Best Fits}

The analysis, as described above, was used to test simple $SO(10)$ models
\cite{adhrs}. It was found \cite{ga_newPRD,ga_newPL} that one of the models,
called model 4c \cite{lucas}, yields very good fits in a large portion of 
the allowed SUSY parameter space. The quality of these fits is presented
in figures 1a-c. The figures show the contour plots of 
the minimum $\chi^2$ in the $(m_0,M_{1/2})$ plane, for three different 
fixed values of the parameter $\mu(M_Z)=80,\,160,\, 240$ GeV.
\footnote
{The figures are taken from \cite{ga_newPRD}. 
These are slightly modified compared to 
the figures in \cite{ga_newPL} due to a
sign error found later in the numerical code. 
}
All initial parameters other than \{$m_0,M_{1/2},\mu$\} were subject 
to minimization. 

Note that as $\mu$ increases, the quality of the fit gets worse. This is 
understood from the form of the SUSY corrections to fermion masses and 
mixings. These corrections increase with $\mu$ and as
$\mu$ gets larger they can only be kept under control by larger squark masses. 
For this reason the contour lines of constant $\chi^2$ move towards larger
values of $m_0$ for $M_{1/2}<400$GeV in figures 1b and 1c.
Varying $\mu$ freely actually results in its approaching 
the lowest possible value. This lower bound on $\mu(M_Z)$ is determined by 
the chargino mass limit from direct searches and is correlated with $M_{1/2}$. 
When the value of $\mu$ is fixed, as in figures 1a-c, the chargino mass limit 
then sets a sharp lower bound on $M_{1/2}$, which is explicitly visible 
in each of the figures. Plots in figures 1a-c were constructed under the 
assumption that the chargino mass limit was 65 GeV.

Because the chargino mass limit has been raised at LEP2, and because
the optimization ends up with low values of $\mu$, in the rest of the 
analysis presented in this paper $\mu(M_Z)$ has been fixed to $110$GeV. 
At this value of $\mu$, the lightest chargino mass turns out to be about 
$100$GeV for $M_{1/2}>340$GeV, and slowly drops down to about $85$GeV for 
$M_{1/2}=200$GeV.
\footnote
{The current limit is $m_{\chi^-}>87$-$90$GeV 
(dependent on tan$\beta$ and the rest of the
SUSY spectrum) from 
the LEP2 run at $\sqrt{s}=183$GeV. \cite{m_C} 
We would need to increase our value of $\mu$ 
by a few GeV in order to get over this limit 
in the region where $M_{1/2}=200$-$225$GeV. 
Such a change would, however, be insignificant 
for the rest of our results.
}
Figures 2a and 2b show explicitly that the structure observed in figures 1a-c 
originates from the two distinct fits corresponding to two separate minima
of the global analysis. The fits are primarily distinguished by the sign
of the effective $\bsgam$ decay amplitude, or in other words, by the sign
of the $C_7$ coefficient of the effective Hamiltonian for the low energy FCNC
processes. The two options are available because of the destructive
chargino interference among the partial amplitudes to this process.
A more detailed discussion of these results 
is actually the subject of section 4.
As can be seen from the contour lines corresponding to $\chi^2=$0.4 and 0.3, 
the minimum in the $(m_0,M_{1/2})$ plane is quite shallow. For this
reason and because the optimization converges very slowly we do not 
specify the exact point with minimum $\chi^2$ in the figures. We are also
aware that such low $\chi^2$ values most likely result from an overestimate
of the theoretical uncertainties. That suggests that our best fit $\chi^2$'s
should be used more in the sense of figures of merit than in a rigorous 
statistical fit evaluation.

We do not show results for negative values of $\mu$. In this case,
the SUSY corrections to $m_b$ are negative (which is rather a welcome 
feature in connection with strict $b-\tau$ unification).
However, the chargino contribution to $\bsgam$ interferes constructively with 
the already large enough SM and charged Higgs contributions. As a result, 
the fits get much worse, with $\chi^2$ well above 
10 per 3 d.o.f., and that disfavors this region of the SUSY parameter space.
Similar observations were also made in \cite{deBoer_Jerus}. 

\section{MSSM Analysis}
\subsection{Model Dependence}

Recall that a general, model independent analysis of the MSSM 
constrained by unification traditionally assumes
exact gauge and Yukawa coupling unification and some degree of universality
among the SUSY mass parameters. The lighter generation 
fermion masses and the CKM matrix elements are taken over from experiment, 
and in the SUSY sector, only the left-right mixings of the stops, sbottoms and 
staus are usually considered, with the inter-generational mixings left out.

The most significant model dependent feature of the analysis presented in 
this letter is the introduction of $\epsilon_3$, the GUT threshold to $\as$. 
We have introduced $\epsilon_3$ as an initial parameter which is free 
to vary within $\pm6$\%. The contour plots of constant $\epsilon_3$ 
resulting from the best fits are shown in the figures discussed 
in section \ref{ss_as}. As one can see, $\epsilon_3$ tends to be negative. 
Lucas and Raby showed that such negative threshold corrections are 
consistent with the complete $SO(10)$ formulation of model 4c \cite{lucas}.
Negative $\epsilon_3$ allows $\as(M_Z)$ to go below 0.120. That in 
turn is a welcome feature for the $b-\tau$ unification if one studies
the SUSY parameter subspace in which $m_b$ receives positive SUSY corrections.
There is more discussion on these effects in section \ref{ss_as}.
$\epsilon_3$ is the only GUT threshold introduced in this study. 

For the Yukawa matrices, the exact equality of the 33 elements is assumed.
The remaining Yukawa 
entries are, of course, dependent on specific properties of model 4c.
However, they are 
small and decouple from the MSSM RGEs for the gauge and third generation 
Yukawa couplings, as well as for the diagonal SUSY mass parameters. 
Thus they have no effect on the calculation of the $Z$-scale values 
for the first nine observables in table \ref{t_obs}.
Only the branching ratio $BR(\bsgam)$ is affected by some of these entries. 
This dependence comes dominantly from the diagram in figure \ref{f_dg}e.
The partial contribution to $\bsgam$ is proportional to the flavor changing 
$\tilde{c}_L$-$\tilde{t}_L$ squark mixing in this case \cite{rosiek} 
and is tan$\beta$ enhanced which makes it non-negligable.
The mixing is completely induced by the off-diagonal 
entries of the Yukawa matrices in the RG evolution, since we assume 
universal squark masses at the GUT scale.
In addition, 
there is no significant pull in the model 4c best $\chi^2$'s from
the light fermion masses and CKM mixing elements \cite{ga_newPRD}. 
Instead, the dominant pulls are imposed by $\as(M_Z)$, $m_b(M_b)$
and other observables which also enter the MSSM analysis. 
Typically, about 60-80\% of the total
$\chi^2$ value comes from the first ten observables of table \ref{t_obs}.
That means
that it is these traditional MSSM observables which drive the model 4c 
optimization procedure when it gets close to its minima 
and that at the same time the Yukawa sector of model 4c works indeed 
very well and does not bias the optimization significantly.
We conclude that 
our results presented in this section are not sensitive to the structure 
of the Yukawa matrices except for the model dependent 23 mixing which 
is significant for the BR($\bsgam$).

\subsection{$BR({\bsgam})$ in MSSM with large tan$\beta$}

The SM, charged Higgs, and chargino diagrams 
contributing to the amplitude for $\bsgam$ are shown in figures 
\ref{f_dg}a--e. These are the dominant contributions
in the best fits of our analysis. Note that the three chargino contributions 
in figures \ref{f_dg}c--e are enhanced by tan$\beta$. 
In our code, we also take into account the gluino and neutralino diagrams 
as well as the full chargino contribution including the pieces 
not enhanced by tan$\beta$, although their contributions are 
numerically insignificant. 

To understand our results which will be presented below, we would like to look 
first at an estimate what to expect from the SUSY contribution to $C_7(M_Z)$. 
As a starting point we acknowledge the fact that the SM contribution gives 
a rough agreement with the measured rate and define the ratios for Higgs
and chargino contributions
\bea
  r^{(H)}    &=& C_7^{(H)}  \: / \:  C_7^{(SM)} \:, \label{rH} \\ 
  r^{(C)}    &=& C_7^{(C)}  \: / \:  C_7^{(SM)} \:, \label{rC}
\eea
where all quantities are evaluated at $M_Z$. 
Equation (\ref{C_7_eff}) then reads
\be
  C_7^{(MSSM)\:eff} \approx \eta^{16\over23}\:C_7^{(SM)}\,
                                       (1+ r^{(H)} + r^{(C)}) 
                      +\sum_{i=1}^8h_i\eta^{a_i}\,
\label{C_7_eff_MSSM}
\ee
where we used $C_2(M_Z)=1$ and, for simplicity, neglected the mixing with
the chromomagnetic operator, proportional to $C_8$, which turns out to be
by about a factor 5--10 less than the other two terms. 

Since $ C_7^{(SM)\:eff} $ (approximately equal to 
$\eta^{16\over23}\,C_7^{(SM)} + \sum h_i\eta^{a_i}\,$)
would yield about the right value of the $BR(\bsgam)$ we deduce that either
\be
   r^{(C)} \approx - r^{(H)} 
\label{r_C_ac}
\ee
for $C_7^{(MSSM)\:eff} \approx + C_7^{(SM)\:eff}$ , or
\be
   r^{(C)} \approx - r^{(H)} - 4.60
\label{r_C_AC}
\ee
for $C_7^{(MSSM)\:eff} \approx - C_7^{(SM)\:eff}$. For the last 
estimate, the numerical results $C_7^{(SM)}=-0.190$, $\eta^{16\over23}=0.679$,
and  $\sum_{i=1}^8h_i\eta^{a_i}\,= -0.168$ were used --- computed for 
$\as(M_Z)=0.118$. 
 
The charged Higgs contribution always interferes constructively with the SM 
contribution \cite{bsgam_theory_largetb}. Typically, we get 
\be
   0 < r^{(H)} < 1.3\,,
\label{r_H}
\ee
depending on the mass of the $H^-$. In the first case, especially  
if $r^{(H)}$ and $r^{(C)}$ are non-negligible, eq.(\ref{r_C_ac})
means that the chargino 
part must interfere destructively\footnote
{Since eq's (\ref{C_7_eff_MSSM}) 
and (\ref{r_C_ac}) are valid only 
approximately, the case 
$1+ r^{(H)} + r^{(C)}\approx 1$
in principle also allows for a 
constructive interference 
$0 < r^{(C)},\: r^{(H)} \ll 1$
in the region in parameter space 
where $m_{H^-}$ and sparticle masses
are large. This option, however, 
does not result from our best fits, 
as already mentioned in the 
discussion on negative $\mu$ 
parameter in the previous section. 
}
with the SM and charged Higgs contributions,
practically cancelling the latter. The enhancement by tan$\beta$ of the 
chargino contribution has to be compensated for by rather large masses of 
the sparticles in the diagrams in figures \ref{f_dg}c--e.
In the second case, described by eq.(\ref{r_C_AC}), large destructive
chargino interference is required to outweigh the combined SM and $H^-$
contributions and to flip the overall sign of the amplitude. 
Quite amazingly, it is not so difficult to arrange (see also \cite{GNg}) 
since the chargino contribution is the only one enhanced by
large tan$\beta$. However, large sparticle masses obviously suppress
the effect. The lesson is that we can expect the two cases to work in
a complementary SUSY parameter space and have 
\sign$\,C_7^{(MSSM)} = 
    \stackrel{+}{\scriptstyle (-)}$ \sign$\,C_7^{(SM)}$ 
for the best fits in the region with large (low) $m_0$ and/or $M_{1/2}$, 
respectively.

These expectations are indeed realized in the best fits of model 4c.
Figure 2a shows the best fits in the case when the chargino contribution
roughly cancels the charged Higgs contribution, while the best fits of 
figure 2b correspond to the case when the chargino piece truly dominates and 
reverses the sign of the overall amplitude. As anticipated, these 
two cases work in complementary regions of parameter space.

Figures \ref{f_BRgam}a and \ref{f_BRgam}b show how well the fits describe 
the measured value of the $BR(\bsgam)$. In these figures (and similarly in the
following ones) we show the contour lines of constant $\chi^2$ in the 
background for reference. As can be seen, 
the $BR(\bsgam)$ indeed presents a major constraint since whenever the 
$\chi^2$ values go up, the agreement with the observed $\bsgam$ decay
rate gets worse. 
Figures \ref{f_rH}a--b and \ref{f_rC}a--b 
show the contour plots of constant $r^{(H)}$ and $r^{(C)}$ 
in each of these two cases. One can compare the numerical results in 
these figures with the approximate relations (\ref{r_C_ac}) and (\ref{r_C_AC}).
Note also the validity of relation (\ref{r_H}) in each case.
Finally note that due to large tan$\beta$ the effects of SUSY decoupling
start showing up only for $m_0>2$TeV, {\em i.e.} outside the SUSY space
studied in these figures.

\subsection{Discussion of the results for $BR({\bsgam})$ and phenomenological
            implications}
There is one striking feature which is common to both cases. It is that both 
fits would like to have the $BR(\bsgam)$ below rather than above the current 
experimental value $2.32\times 10^{-4}$. 

In the first case, the tan$\beta$
enhanced chargino contribution tends to be too large when going against 
the charged Higgs contribution, since the latter is not tan$\beta$ enhanced.
The fit clearly favors as large Higgs contribution as possible with $r^{(H)}$
reaching its maximum (see fig.\ref{f_rH}a).
A phenomenological consequence of this observation is that the charged Higgs 
(and then also the whole Higgs sector) tends to be as light as possible 
in this case. We get, for instance, the best fit value of the pseudoscalar 
mass $m_A<100$GeV everywhere in the $(m_0,M_{1/2})$ plane.  

In the second case, when the chargino part overshoots the combined SM
and $H^-$ contributions we observe different effects in the regions
with $M_{1/2}$ below and above (roughly) $300$GeV.
For larger values of $M_{1/2}$ the chargino contribution clearly
tends to be not big enough. As a result we might expect to see
only very low values of $r^{(H)}$ in this region --- complementary to 
the large values in fig.\ref{f_rH}a --- and a very heavy Higgs sector.
However, the best fit value of $r^{(H)}$ varies quite a bit indicating 
that the charged Higgs mass does not stay at some very large value. 
That is related to the observation \cite{ga_newPRD} that one 
cannot have good fits with the Higgs sector much heavier than squarks. 
When $M_{1/2}$ gets below $300$GeV 
the $\bsgam$ decay rate is no longer a strong constraint and two separate 
minima can be found in the course of the optimization. The two fits work
equally well: $\chi^2$ in each case stays below 1 per 3dof. The minima 
differ by the 
best fit values of the Higgs masses: one minimum corresponds to $m_{h^0}$ 
and $m_{A}$ at the experimental lower limit (set to $65$GeV in our analysis) 
while these masses gradually rise in the second ``valley'' up to $700$GeV. 
When crossing the region with $M_{1/2}\approx
300$GeV towards larger $M_{1/2}$, the first ``valley'' vanishes and 
the optimization slides down to the second minimum because of the $BR(\bsgam)$.
In the 
allowed corner with $m_0<700$GeV, the two ``valleys'' approach each 
other and finally coincide.
The effect of the doubled minima is indicated in figures 
\ref{f_rH}b and \ref{f_rC}b with the solid black (dashed gray) contour lines 
corresponding to the heavier (lighter) Higgs sector\footnote
{The same holds in figures 
\ref{f_rC23}b and \ref{f_dmb}b.
}.
  
In summary, if the future experimental analysis confirms the discrepancy
between the CLEO measured value and the NLO SM calculation \cite{NLO_SM},
the MSSM with large tan$\beta$ could be the solution. Similarly, if
the NLO calculation is completed for the MSSM, and if it turns out to
increase the LO result as occurred for the SM,
then the large tan$\beta$ regime will apparently have no problem fitting
the $\bsgam$ rate exactly. That seems to be in contrast with the fits
in the low tan$\beta$ regime of ref.\cite{deBoer_Jerus}, which get 
below the SM value only for $M_{1/2}<200$GeV and small $m_0$.

\subsection{Role of $\tilde{c}$-$\tilde{t}$ mixing in our results for $BR({\bsgam})$}
It is interesting to note the significance of the 
inter-generational squark mixing in fig.\ref{f_dg}e. In analogy to
eq's (\ref{rH}), (\ref{rC}) we define
\be
  r^{(C23)} = C_7^{(C23)} \: / \:  C_7^{(SM)} \:, 
\label{rC23}
\ee
where $C_7^{(C23)}$ is the $\tilde{c}_L$-$\tilde{t}_L$ mixing contribution 
to the coefficient $C_7$ at the scale $M_Z$. We show the contour plots 
of the constant $r^{(C23)}$ in figures \ref{f_rC23}a and \ref{f_rC23}b. 
While the dominant chargino contribution is that proportional to the 
$\tilde{t}_L$-$\tilde{t}_R$ mixing (fig.3c), $C_7^{(C23)}$ becomes important 
because of the destructive character of the interference among the partial 
amplitudes. As one can see, the $\tilde{c}_L$-$\tilde{t}_L$ mixing term can 
be comparable with the SM and $H^-$ contributions. More importantly, it 
always interferes constructively with them. In the case when 
\sign$\,C_7^{(MSSM)} =  +$ \sign$\,C_7^{(SM)}$, this term
helps to counterbalance the large contribution of the left-right stop
mixing. In the complementary case,
when the chargino contribution overturns the sign
of the net amplitude the $\tilde{c}_L$-$\tilde{t}_L$ mixing makes this flip
more difficult to happen. As a result, it has different consequences for
the two fits in figures 2a and 2b, especially important  
in the region ($m_0\leq1000$GeV,$\,M_{1/2}\approx350$GeV) where the fits 
start getting worse. In figure 2a, it improves the fit for lower values of 
$(m_0,M_{1/2})$ where the contribution from the stop mixing alone would 
otherwise overwhelm the sum of the SM and $H^-$ diagrams. On the contrary, 
it worsens the fit in figure 2b in the same parameter subspace.
These observations are, however, model dependent as has been explained 
in the introduction to this section, and their validity relies
on the boundary conditions assumed at the GUT scale. A general
study of the limits imposed by the inter-generational mixings can be found 
in the review \cite{rosiek} and the references therein. To make a connection
with the general approach we note that typically we get 
\be
 (\delta^{23}_U)_{LL}\equiv \frac{(m^2_U)^{23}_{LL}}
                                 {[(m^2_U)^{22}_{LL}(m^2_U)^{33}_{LL}]^{1/2}}
                     \sim -(0.01\mbox{-}0.02)\,
\label{delta}
\ee
at the $Z$ scale, where $(m^2_U)_{LL}$ is the $3\times3$ $m^2_Q$ squark mass 
matrix after the unitary rotations which diagonalize the fermionic sector are
performed in the squark sector too ( --- sandwiching by $V_U^L$'s in this 
particular case). 
This value is to be compared with the flavor--changing effects originating 
in the CKM matrix (figures \ref{f_dg}a--d) with 
$|V_{ts}|\approx|V_{cb}|\approx 0.04$. 

\subsection{Effects of NLO QCD Corrections to BR($\bsgam$)}
\label{ss_NLO}
Finally, we would like to comment on the next-to-leading order (NLO) QCD
corrections. These have been completed only for the SM \cite{NLO_SM}. 
In ref.\cite{ay} it was shown that the NLO matching at the 
high scale gives a non-negligible contribution to $C_7$. In particular,
for the SM we get 
\bea
  C_7^{eff}(\mu_b)&=&C_7^{(0)\,eff}(\mu_b) 
   + \frac{\as}{4\pi}\, 
    (\, \eta^{39/23}C_7^{(1)}(M_Z) 
     +\frac{8}{3}(\eta^{\frac{37}{23}}-\eta^{\frac{39}{23}}) C_8^{(1)}(M_Z) 
                       + \cdots) \nonumber \\ 
  \mbox{ } &=& -0.312 + 0.017\,(-0.883 - 0.111 + \cdots)\;,
\label{C_7_NLO}
\eea
where $C_7^{(0)\,eff}(\mu_b)$ is given in terms of the LO matching 
coefficients (our eq.(\ref{C_7_eff})) and for the evaluation
in the second line $\as(M_Z)=0.118$ was assumed.
Hence within the SM the NLO matching corrections $C_7^{(1)}(M_Z)$ and 
$C_8^{(1)}(M_Z)$ alone would change the rate by about 10\%.
The final NLO corrected result increases the
LO value obtained at $\mu_b\simeq M_b$ by about 20\%. 
(In other words,
it effectively lowers the scale $\mu_b$ of the LO calculations to about 
$0.6\,M_b$ as obtained also in ref.\cite{deBoer_Jerus}.) 
In the MSSM, the NLO matching remains to be calculated. Clearly, with 
large tan$\beta$ the chargino diagrams will likely dominate and their 
contribution may cause a different effect than in the SM. For that reason 
we have left our results without applying the NLO corrections 
and will just sketch at this point what effects the higher order matching 
corrections may have in our best fits.

The case when \sign$\,C_7^{(MSSM)} =  +$ \sign$\,C_7^{(SM)}$ is much like
the SM case. The effect of the unknown $C_7^{(MSSM)\,(1)}(M_Z)$ and 
$C_8^{(MSSM)\,(1)}(M_Z)$ can be readily estimated from the complete
form of eq.(\ref{C_7_NLO}) quoted in \cite{NLO_SM}. 
If the NLO matching followed the cancellation among the charged Higgs and 
chargino contributions observed in the LO (eq.(\ref{r_C_ac})), our results 
for the $BR(\bsgam)$
would also be about 20\% larger, similar to the SM calculation. 
We can see from figure \ref{f_BRgam}a 
that this would increase the parameter space in this case. Effectively, 
the NLO terms would add to the LO contributions of $W^-$ and $H^-$,
enabling the tan$\beta$ enhanced chargino contribution to be greater 
in magnitude than what is allowed by eq.(\ref{r_C_ac}) --- a welcome 
feature for lighter SUSY spectra. Clearly, if the NLO matching terms remain 
negative and get even larger in magnitude than in the SM case, the fit will 
improve substantially in the region with low $m_0$ and $M_{1/2}$ (and  
may get somehow worse in the upper left corner of the $(m_0,M_{1/2})$ plane). 
On the other hand, we have checked that our LO results remain unaltered for 
$C_k^{(MSSM)\,(1)} \simeq - C_k^{(SM)\,(1)}$, $k=7,8$: for such matching terms 
all NLO corrections practically sum up to zero. 
Only if the NLO matching terms turn out to be very 
large and positive, the fit will be forced to retreat significantly towards
larger $M_{1/2}$ values. 

The second case, \sign$\,C_7^{(MSSM)} =  -$ \sign$\,C_7^{(SM)}$, is obviously 
more sensitive to the unknown NLO matching. Despite the sensitivity our sample 
calculations indicate that a drop in the value of the $BR(\bsgam)$ is fairly 
common and more likely to happen than an enhancement. For instance, we get 
$BR(\bsgam)\times10^4 = 0.98\, (2.31)$ in the complete NLO calculation
assuming the NLO matching terms are 
$C_k^{(MSSM)\,(1)} = +1\, (-7)\times C_k^{(SM)\,(1)}$, $k=7,8$, 
and keeping $C_7^{(0)}$ and $C_8^{(0)}$ at the values which yield 
$BR(\bsgam)\times10^4 = 2.31$ at the LO. 
The reduction in the rate is correlated with the opposite
signs of $C_7$ and $C_8$ as compared to the SM case. However, it still 
holds that the NLO correction would be effectively taken into account by
lowering the scale in the LO calculation. 
For the region with $M_{1/2}>300$GeV, where the fit gradually gets worse
in this case, it means that the $\bsgam$ decay is likely more constraining 
than it appears in the analysis based on the LO calculation. As indicated 
above, the NLO corrections will improve the fit only if the NLO matching terms 
are at least seven times larger and opposite in sign compared to the SM NLO 
matching terms.

\subsection{$\delta m_b^{SUSY}$, $\as(M_Z)$ and $\epsilon_3$ in the Best Fits}
\label{ss_as}

For the presented results, it has been important to have a destructive 
interference
among the partial contributions to $\bsgam$. With the universal boundary
conditions at the GUT scale however, that can be arranged only for a
specific sign of the $\mu$ parameter: $\mu>0$ in our conventions.
That in turn 
correlates with the positive sign of the SUSY corrections to the $b$
quark mass \cite{copw}. This fact implies strong constraints on the 
SUSY parameter space 
from $m_b$ since $\delta m_b^{SUSY}$ gets a dominant contribution 
from the tan$\beta$ enhanced gluino exchange, which tends to be very large
\cite{hrs}
and offsets the agreement between the low energy value of $m_b$ and
the $b$--$\tau$ unification at the GUT scale. The gluino correction 
can be explicitly suppressed by heavy squarks and by the 
chargino correction which enters with the opposite sign\footnote
{The sign of the chargino induced
correction to $m_b$ is determined 
by the product $A_t\mu$ (the 
dominant part comes from the same 
diagram as in fig.\ref{f_dg}c 
with $s_L$ replaced by $b_L$ and no 
photon leg attached). Our best fits
always run into the region where
$A_t(M_Z)$ turns out negative, 
in the conventions which maintain
positive gluino mass parameter.
}.
It can also be reduced by lower values of $\as(M_Z)$ because of 
the strong couplings in the vertices of the gluino -- $b$-squark diagram.
The contour plots of the total SUSY correction to $m_b(M_Z)$ in the best
fits are shown in figures \ref{f_dmb}a--b where one can see that these 
effects are quite effective in reducing the value of $\delta m_b$.

Note that the impact of large positive 
$\delta m_b^{SUSY}$ is reduced by a lower value of $\as(M_Z)$ 
also indirectly ---  due to the RG evolution of the current $b$ mass 
from $M_Z$ down to $M_b$. 
The effect is enhanced if the current mass $m_b$ is converted to the 
perturbative pole mass $M_b$ in a top--down analysis.
For example, the difference between $\as(M_Z)$ being 0.115 and 0.121
leads to about 5\% difference in $M_b$, if the same value of $m_b(M_Z)$ 
is assumed in each case. This is a significant effect. 
In our analysis, which assumes larger uncertainty for $\as$ than for
$m_b(M_b)$, it pushes $\as(M_Z)$ down. The effect can be seen in figures
\ref{f_alpha_s}a--b. Note that one can trade lower values of $\as$ 
for higher values of $m_b$ provided a smaller uncertainty $\sigma(\as)$
is assumed \cite{PRDpaper}.

Rather low values of $\as(M_Z)$ are, however, difficult to obtain from
the exact gauge coupling unification, traditionally assumed in
an MSSM analysis constrained by unification. In our analysis, we assumed
that there is a few per cent correction to the gauge coupling unification
generated by the spread in masses of heavy states integrated out at
the GUT scale. It turned out that with a few per cent negative correction
$\epsilon_3$ to $\as(M_G)$ we can accommodate lower values of $\as(M_Z)$
with no problem. The best fit values of $\epsilon_3$ are presented 
in figures \ref{f_eps3}a--b. Clearly, as squarks get lighter
the SUSY correction to $m_b(M_Z)$ (fig.\ref{f_dmb}) has to be increasingly 
reduced by lower values of $\as(M_Z)$ (fig.\ref{f_alpha_s}), which in 
turn requires a more substantial departure from the gauge coupling unification 
(fig.\ref{f_eps3}) in the respective SUSY parameter subspace.

\section{Conclusions}
In summary, we have presented the results of the MSSM analysis
focusing on the constraint imposed by the measured value of the
$BR(\bsgam)$. The analysis was motivated by the best of the $SO(10)$ SUSY GUT 
models studied previously. Large tan$\beta$ was assumed as a consequence of 
simple SO(10) flavor dynamics. We showed that the effective amplitude 
for the inclusive $\bsgam$ decay can be of either sign in such a scenario.
In the case, when the sign is the same as in the SM calculation, the
often neglected inter-generational $\tilde{c}_L$-$\tilde{t}_L$ squark mixing 
and the NLO correction tend to increase the allowed SUSY parameter space.
The best fits favor light Higgs spectrum since a substantial charged Higgs 
contribution to $\bsgam$ helps to counterbalance the chargino contribution.  
In the complementary case, when the sign of the $C_7$ coefficient is flipped
by a large chargino contribution, both the $\tilde{c}_L$-$\tilde{t}_L$ mixing 
and the NLO correction work the other way and tend to reduce the 
allowed SUSY parameter space. The charged and pseudoscalar Higgs masses can be 
either large or small. In each case, the chargino (and neutralino) masses are
at the experimental limit in the best fits. Because of $\delta m_b^{SUSY} >0$
the best fits result in 
rather low values of $\as(M_Z)$, which can be obtained with a few per cent 
negative correction to $\as(M_G)$. With these guidelines the large tan$\beta$
regime of the MSSM constrained by simple $SO(10)$ GUTs remains a viable
option for physics beyond the Standard Model.

\vskip 1cm
\noindent {\Large\bf Acknowledgments}
\vskip 0.5cm
The authors would like to thank Marcela Carena
and Carlos Wagner who collaborated on the early stages of this project. 
This research was supported in part by the U.S. Department of Energy under
contract numbers DE-FG02-91ER40661 and DOE/ER/01545-729.

\newpage

\begin{figure}[p]
\epsfysize=7.0truein
\epsffile{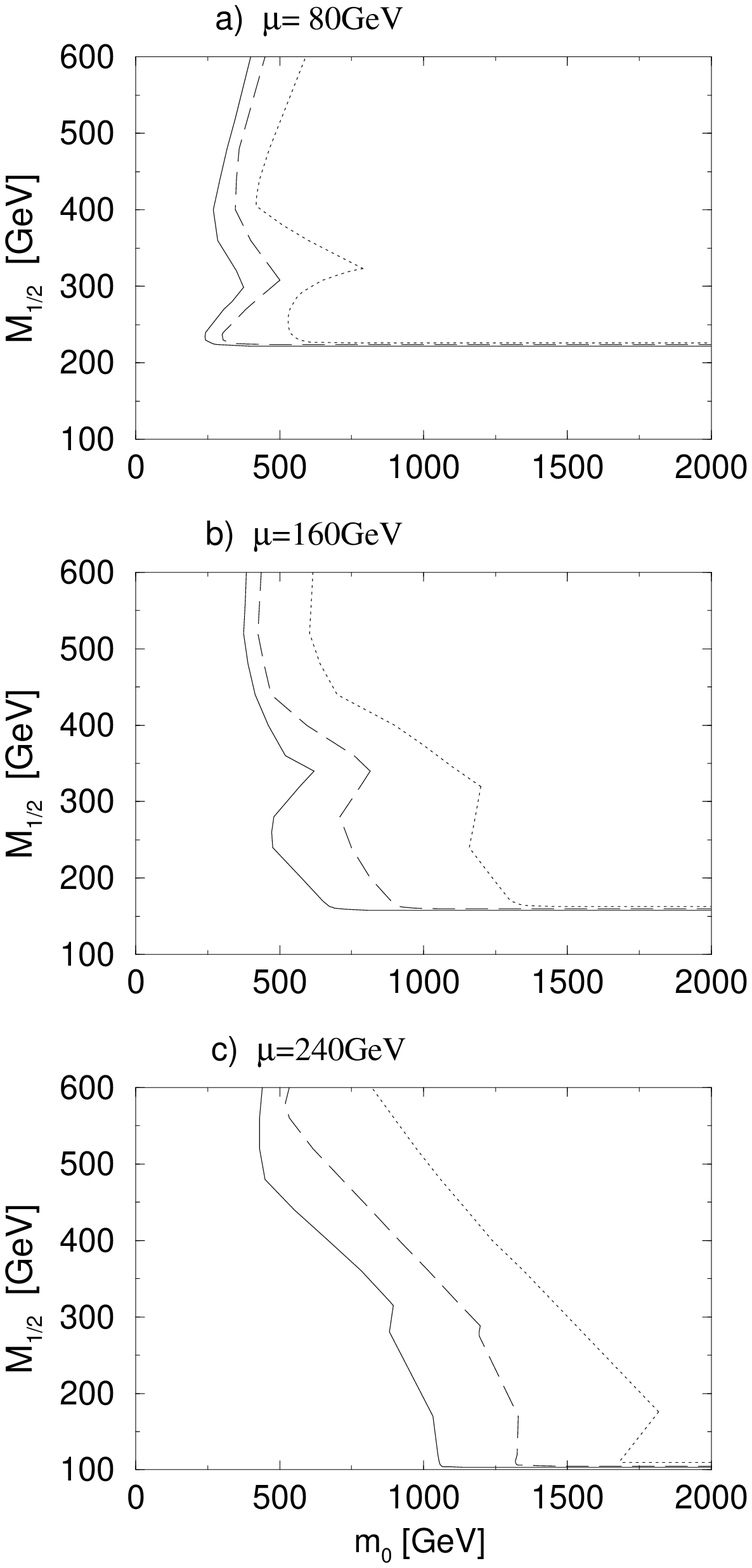}
\caption{
Contour lines of constant $\chi^2$ in the best fits of model 4c
for $\mu(M_Z)$ = ($a$) 80 GeV,  ($b$) 160 GeV and ($c$) 240 GeV.
Solid, dashed and dotted curves correspond to $\chi^2$= 6, 3 and 1 
{\em per} 3 degrees of freedom, respectively.
}
\label{f_F4c}
\end{figure}

\begin{figure}[p]
\epsfysize=7.0truein
\epsffile{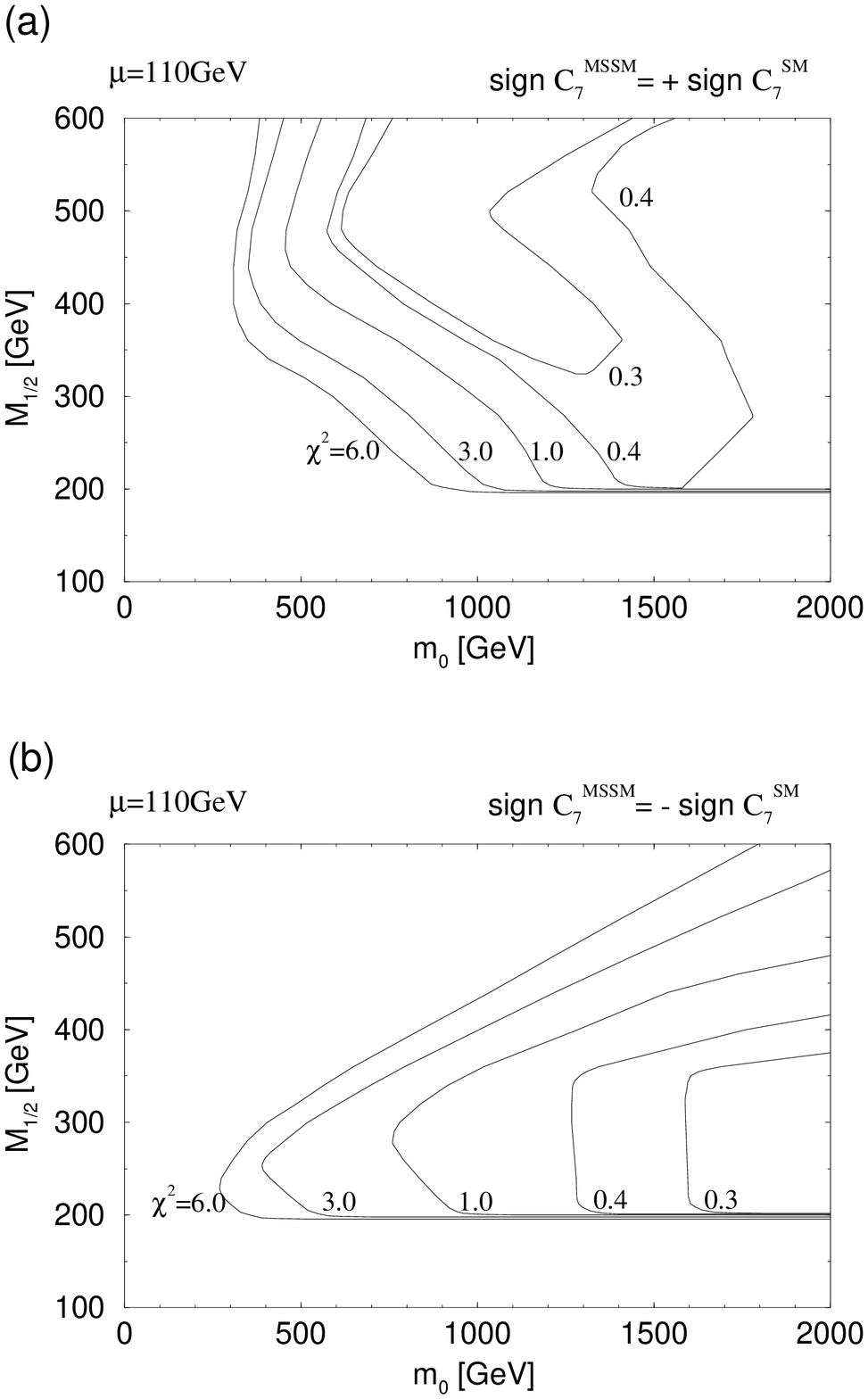}
\caption{
$\chi^2$ contour plots in the best fits of model 4c 
with the MSSM effective amplitude for $\bsgam$ of ({\em a}) the
same ({\em b}) the opposite sign as compared to the SM amplitude. 
$\mu(M_Z)$ = 110 GeV, which --- as described in the text ---    
basically corresponds to the minimum value if $\mu$ freely 
varied as an input parameter. As indicated, curves correspond
to $\chi^2=6,\,3,\,1,\,0.4,\,0.3$ per $3\:d.o.f.$, respectively.
}
\label{f_F4c110}
\end{figure}

\begin{figure}[p]
\epsfxsize=5.5truein
\epsffile{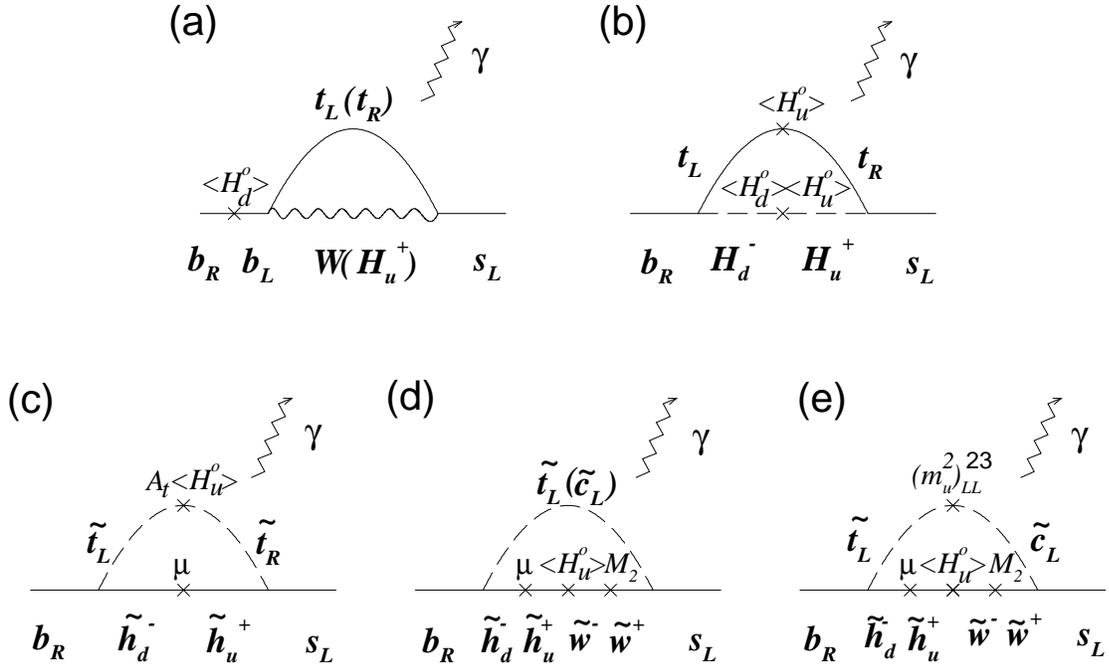}
\caption{
The most important diagrams for the $\bsgam$ decay in the MSSM with
large tan$\beta$, in the interaction basis. Note the enhancement of
the chargino diagrams (c)--(e) by tan$\beta$ compared to the 
SM and charged Higgs diagrams (a) and (b). 
}
\label{f_dg}
\end{figure}

\begin{figure}[p]
\epsfysize=7truein
\epsffile{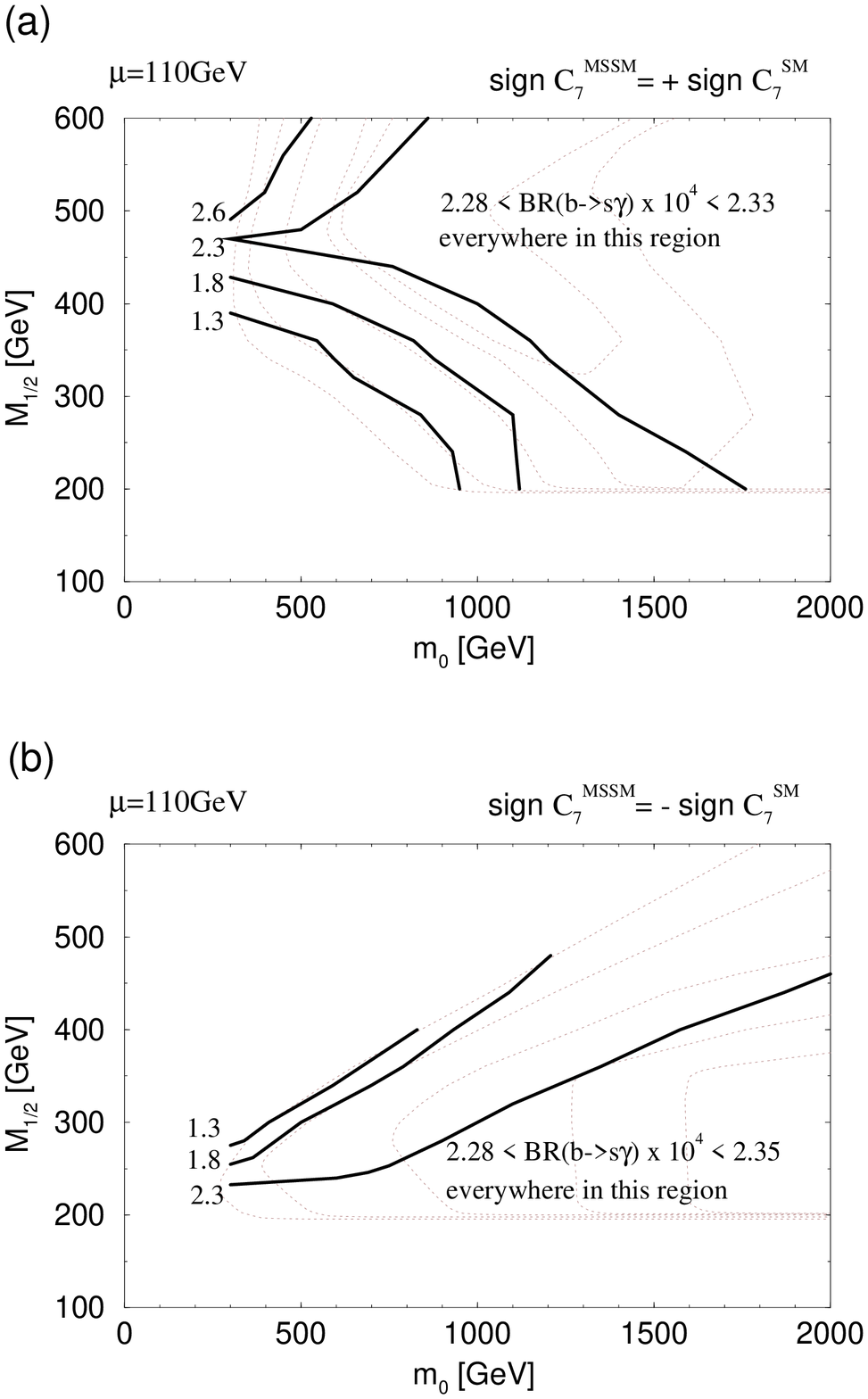}
\caption{
Contour lines of constant $BR(\bsgam)\times 10^4$ in the best fits 
assuming the opposite signs of the effective amplitude for this process.
For better reference, the $\chi^2$ contour plots of figures 2a and 2b
are shown in the background of (a) and (b), respectively, as dotted lines.
}
\label{f_BRgam}
\end{figure}

\begin{figure}[p]
\epsfysize=7truein
\epsffile{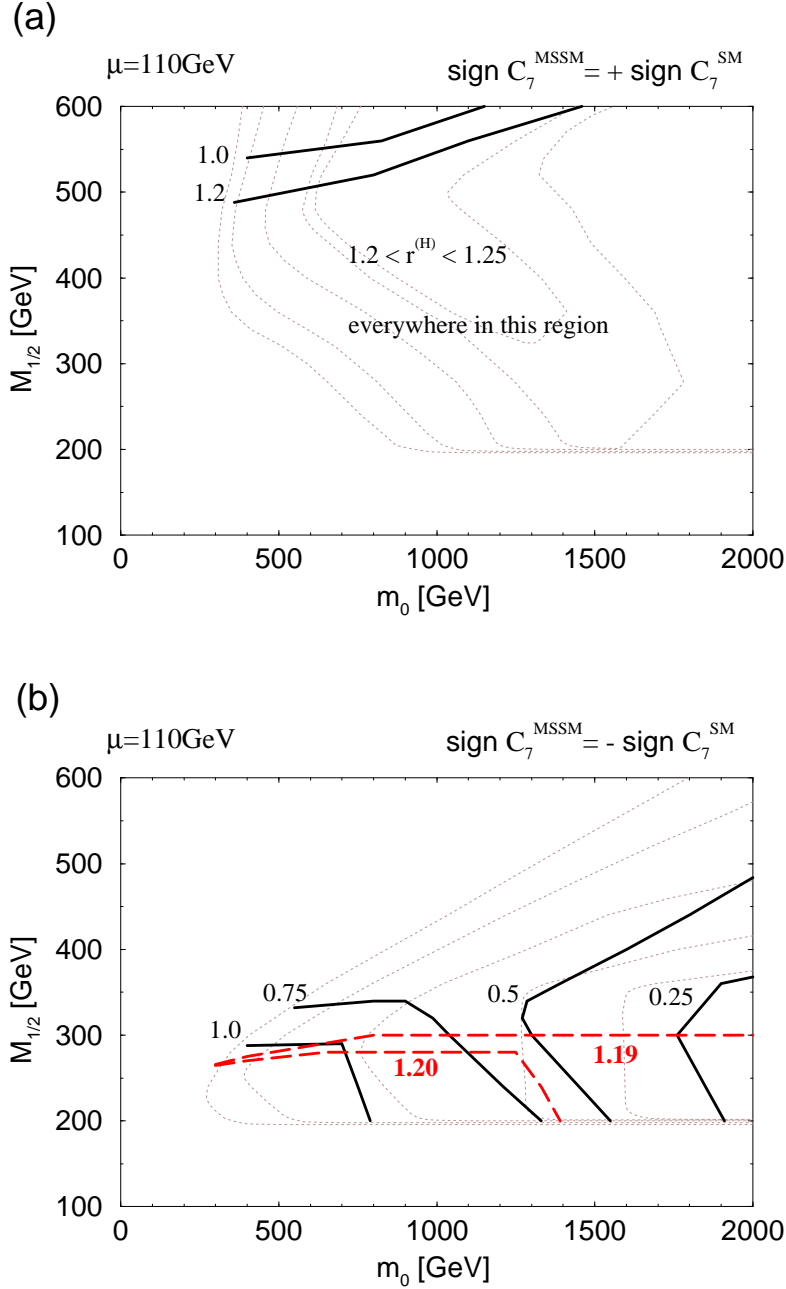}
\caption{
The same as in fig.\ref{f_BRgam} for the contour plot of 
$r^{(H)}$ defined in eq.(\ref{rH}). The effect of the doubled
minimum in the lower part of figure (b) corresponds to the possibility of
having heavy (solid black curves) or light (dashed gray curves)
Higgs spectrum for $M_{1/2}<300$GeV.
}
\label{f_rH}
\end{figure}

\begin{figure}[p]
\epsfysize=7truein
\epsffile{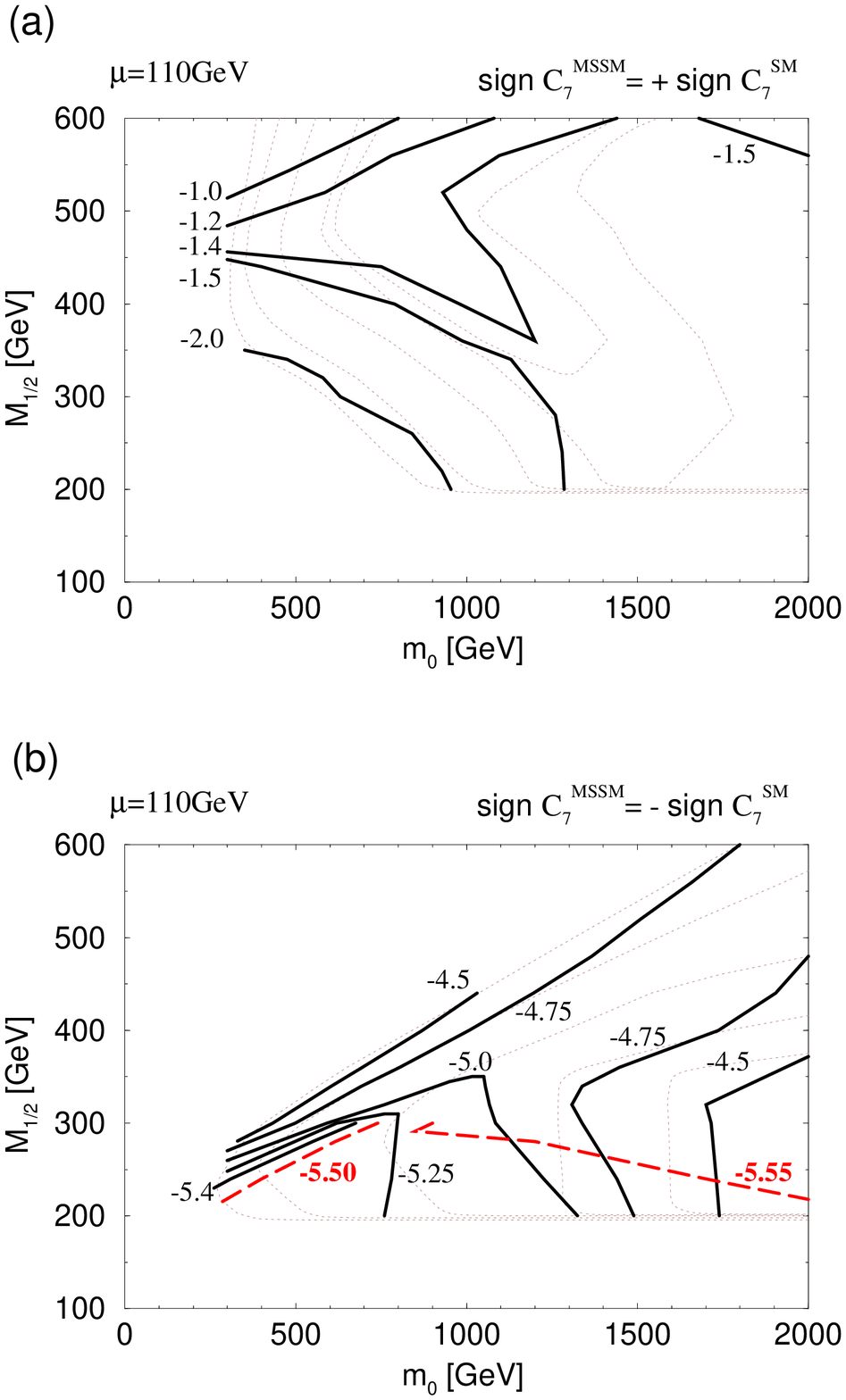}
\caption{
The same as in fig.\ref{f_rH} for the contour plot of 
$r^{(C)}$ defined in eq.(\ref{rC}).
}
\label{f_rC}
\end{figure}

\begin{figure}[p]
\epsfysize=7truein
\epsffile{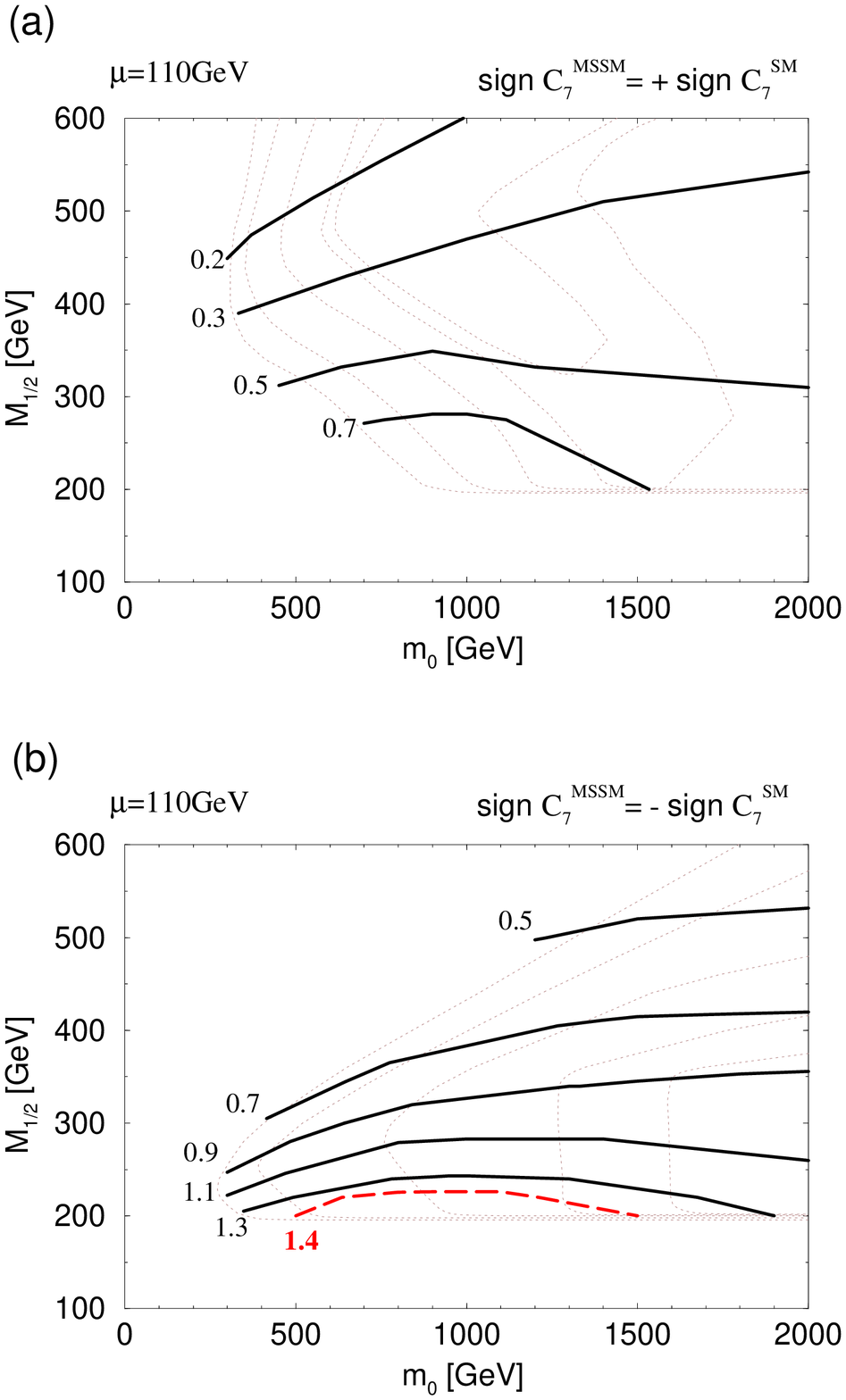}
\caption{
The same as in fig.\ref{f_rH} for the contour plot of 
$r^{(C23)}$ defined in eq.(\ref{rC23}).
}
\label{f_rC23}
\end{figure}

\begin{figure}[p]
\epsfysize=7truein
\epsffile{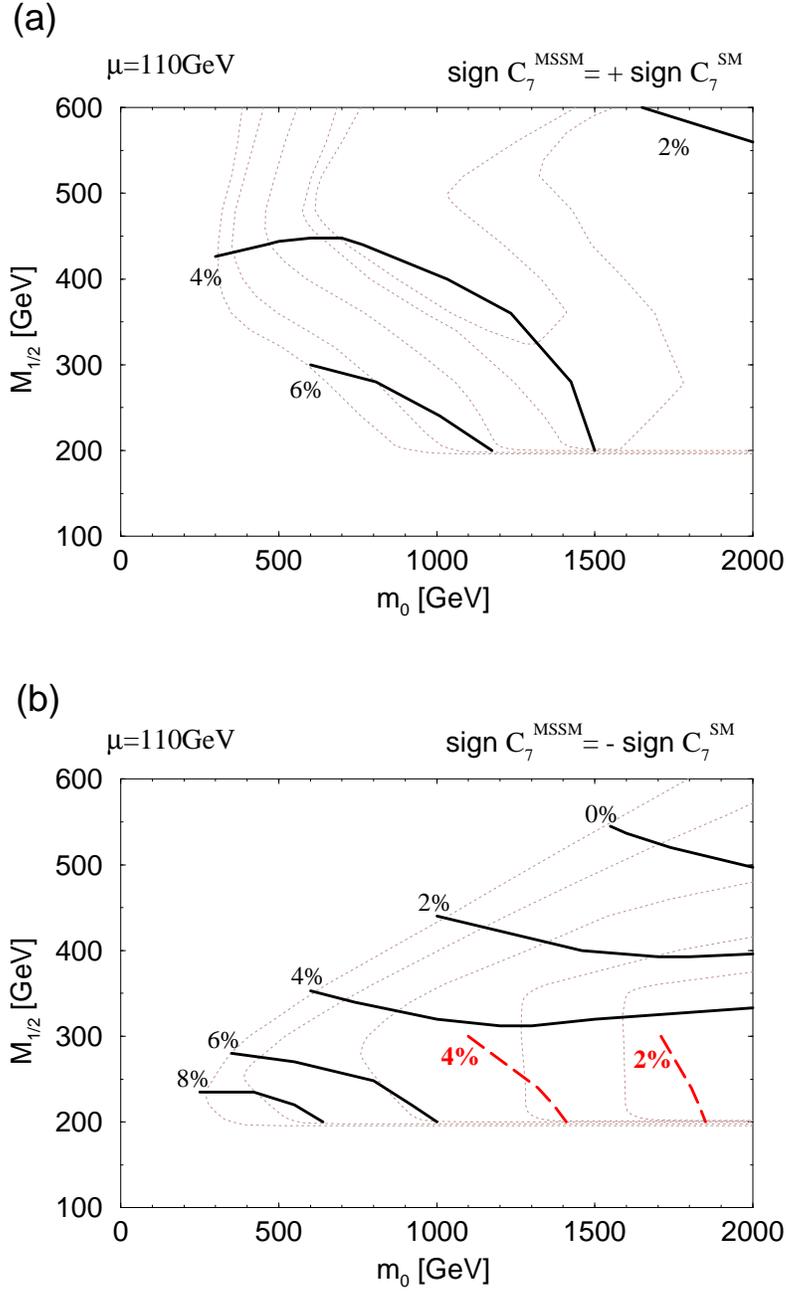}
\caption{
The same as in fig.\ref{f_rH} for the contour plot of 
$\delta m_b^{SUSY}$, the SUSY correction to the $b$ quark mass $m_b(M_Z)$.
}
\label{f_dmb}
\end{figure}

\begin{figure}[p]
\epsfysize=7truein
\epsffile{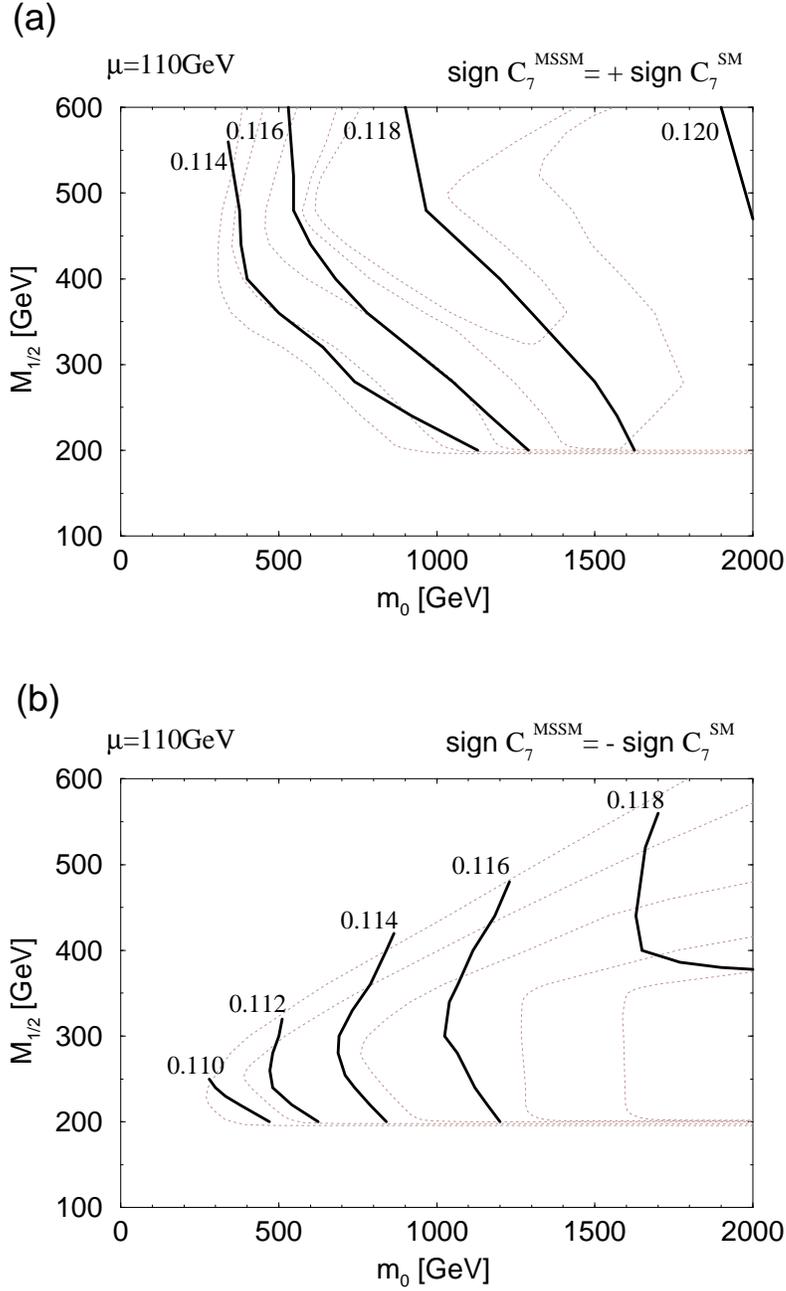}
\caption{
The same as in fig.\ref{f_BRgam} for the contour plot of 
$\as(M_Z)$.
}
\label{f_alpha_s}
\end{figure}

\begin{figure}[p]
\epsfysize=7truein
\epsffile{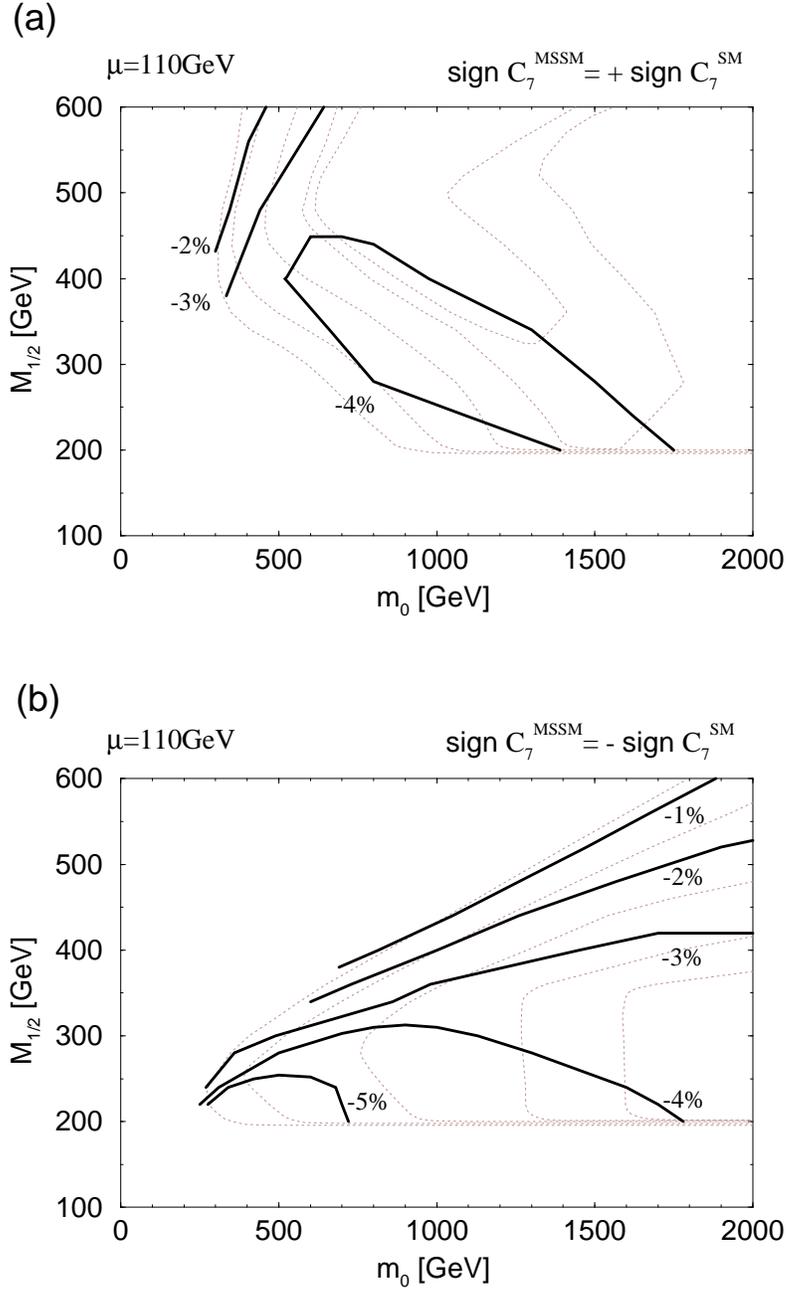}
\caption{
The same as in fig.\ref{f_BRgam} for the contour plot of 
$\epsilon_3$, the GUT scale threshold correction to $\as$.
}
\label{f_eps3}
\end{figure}

\end{document}